\documentclass [pra, twoocolumn, showpacs, superscriptaddress, 10pt] {revtex4-2}%
\usepackage{graphicx}
\usepackage{amsmath}

\begin{document}

\title{Time-ordering effects in a one-atom laser based on electromagnetically-induced transparency}

\author{Dmitri B. Horoshko}\email{horoshko@ifanbel.bas-net.by}
\affiliation {B.~I.~Stepanov Institute of Physics, NASB, Nezavisimosti Ave.~68, Minsk 220072 Belarus}

\author{Chang-shui Yu}
\affiliation {School of Physics, Dalian University of Technology, Dalian 116024, China}

\author{Sergei Ya. Kilin}
\affiliation {B.~I.~Stepanov Institute of Physics, NASB, Nezavisimosti Ave.~68, Minsk 220072 Belarus}

\begin{abstract}
One-atom laser based on electromagnetically-induced transparency, suggested recently [Phys. Rev. Lett. {\bfseries 124}, 093603 (2020)], is capable of generating Schr\"odinger cat states in the regime of strong ground-state coupling. In this regime, we find the exact solution for the Schr\"odinger equation with a time-dependent effective Hamiltonian by considering the Magnus expansion of the time-ordered exponential and calculating analytically the time-ordering terms, omitted in the previous study. We show that the time-ordering term affects the relative phase of two coherent components of the generated Schr\"odinger cat state. We show this influence by calculating various nonclassicality indicators for the cavity field, such as total noise, average parity and relative total noise. We find, that time-ordering becomes important at the average photon number in the cavity below 1, in striking contrast to the case of single-pass parametric downconversion, where it becomes important at average photon number in one optical mode above 4.
\end{abstract}

\maketitle

\section{Introduction}

One atom laser (OAL) is a laser, using just one atom as its active medium, which was shown to be enough for light amplification by stimulated emission of radiation \cite{Mu92}. Realizations of OAL use a neutral atom \cite{Mckeever03} or an ion \cite{Dubin10} trapped in a definite position inside a high-finesse optical cavity. Closely related concepts are one-atom maser \cite{Raimond82,Meschede85}, realizing a similar atom-field interaction in the microwave domain, and single-quantum-dot nanolaser \cite{Strauf11}, where the single emitter is represented by a quantum dot in a semiconductor matrix. All these microscopic laser devices share one feature: the state of the generated cavity field may be highly nonclassical \cite{Pellizzari94,Kilin02,Kilin12,Larionov13,Stefanov19} and therefore a correct description of the atom-field interaction requires field quantization, in contrast to the conventional laser, where a semiclassical description is often sufficient. Studying of these devices is important for better understanding the fundamental laws of interaction of single emitters with quantized electromagnetic field. Besides, the generated nonclassical states of light are prospective for applications in quantum metrology and quantum information processing \cite{Kilin01}.

One of the main limiting factors for producing highly nonclassical optical states in OAL is spontaneous decay from the excited atomic level. Recently, it was suggested \cite{Villas-Boas20} to use the technique of electromagnetically-induced transparency (EIT) \cite{Bermel06,Mucke10,Slodicka10,Kampschulte10,Witthaut10,Xiu16,Mirza18} in OAL, employing an atom in $\Lambda$-configuration with two ground and one excited state. In this configuration, under conditions of EIT, the population of the excited state of the atom is very low, which dramatically decreases the influence of its spontaneous decay on the operation of the laser. Two regimes of the EIT-based OAL (EIT-OAL) are distinguished. In the regime of weak ground-state coupling, which was recently realized in an experiment \cite{Tolazzi21}, OAL works as a single-photon source. In the strong ground-state coupling an elaborated theoretical model predicts generation of Schr\"odinger-cat states of the optical cavity field. We note that the model of Ref.~\cite{Villas-Boas20} disregards the time ordering in the evolution operator, which is a good approximation for a small number of photons in the cavity only. Generation of superpositions of two macroscopically (or at least mesoscopically) distinguishable states, known as optical analogs of the Schr\"odinger cat state, may thus require considering the time-ordering effects in the evolution of a quantum system with a time-dependent Hamiltonian \cite{Quesada15,Horoshko18josab}. Indeed, it was shown for the squeezed states of light both numerically \cite{Christ13} and analytically \cite{Lipfert18} that time-ordering becomes important at squeezing above 12.5 dB, which corresponds, for a squeezed vacuum state, to approximately 4 photons on average in one optical mode. On the other hand, a superposition of two coherent states with opposite phases becomes a Schr\"odnger cat state at the average photon number above 2 \cite{Ourjoumtsev09}.

The main aim of this article is to consider the evolution operator of EIT-OAL in the regime of strong ground-state coupling in a form of Magnus expansion taking into account the time-ordering terms of all orders. Fortunately, the Magnus series breaks off at the second term, allowing us thus to obtain an exact solution for the atom-field dynamics. A secondary aim is to analyze the state of the generated cavity field with the help of various measures of nonclassicality and to observe the effect of time ordering on these measures. In Sec. II we give a general description of the considered system and deduce the effective Hamiltonian and the atom-field state without time-ordering, mainly following Ref.~\cite{Villas-Boas20}. In Sec. III we provide the exact solution for the evolution operator with a time-dependent Hamiltonian and show the effect of time ordering on the state of the generated optical field by calculating various nonclassicality indicators for the cavity field, such as total noise, average parity and relative total noise. Sec. IV concludes the article.

\section{Temporal evolution of atom and field in EIT-OAL}
\subsection{System description in the Schr\"odinger picture}

\begin{figure}[ht!]
\centering\includegraphics[width=6cm]{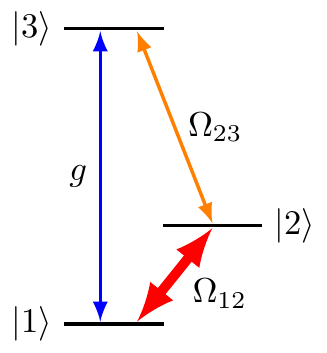}
\caption{Energy levels of EIT-OAL. Two ground states $|1\rangle$ and $|2\rangle$ are connected to the excited state $|3\rangle$ via coupling to optical modes. One of them, in the transition $|1\rangle\leftrightarrow |3\rangle$, is the quantized mode of the cavity field, $g$ being the coupling constant. Another, in the transition $|2\rangle\leftrightarrow |3\rangle$ is a classical coherent wave with an amplitude corresponding to Rabi frequency $2\Omega_{23}$. Two ground states are connected via a
Raman transition with an intermediate state (not shown) pumped by two additional lasers, such that the corresponding Rabi frequency is $2\Omega_{12}$. All fields are assumed to be resonant with the corresponding transitions. \label{fig:levels}}
\end{figure}

The energy levels of EIT-OAL \cite{Villas-Boas20,Tolazzi21} are shown in Fig.~\ref{fig:levels}. In the strong ground-state coupling regime, considered here, we assume $\Omega_{12}\gg(g,\Omega_{23})\gg\kappa$, where $\kappa$ is the cavity-field decay rate. Smallness of the latter with respect to the Rabi frequencies of atomic transitions means that the atom makes many cycles of transitions before a cavity photon is lost. Therefore, one can disregard this loss mechanism for the atom-field evolution at interaction times shorter than $\kappa^{-1}$, which we assume in the following. One can also disregard the spontaneous emission from the exited state because, as we will see later, the population of this level is negligible in the same period of time.

Thus, we describe the atom-field evolution by a unitary evolution operator $\mathcal{U}(t)$, satisfying the Schr\"odinger equation ($\hbar=1$)
\begin{equation}\label{SE}
  i\frac{d}{dt}\mathcal{U}(t) = H \mathcal{U}(t),
\end{equation}
with the Hamiltonian $H = H_0 + V_0$, where
\begin{eqnarray}\label{H0}
  H_0 &=& \sum_{n=1}^3E_n\sigma_{nn} + \omega_{13}a^\dagger a + \omega_{12}b^\dagger b + \omega_{23}c^\dagger c,\\\label{V0}
  V_0 &=& ga\sigma_{31} + d_{12}b\sigma_{21} + d_{23}c\sigma_{32} + \textrm{H.c.}
\end{eqnarray}
Here $E_n$ is the energy of the $n$th atomic level, $\omega_{nm} = E_m-E_n$ is the circular frequency of the field mode resonant with the atomic transition $|n\rangle\rightarrow|m\rangle$, $a$, $b$, and $c$ are the photon annihilation operators of the corresponding optical modes, $\sigma_{nm} = |n\rangle\langle m|$, and $d_{nm}$ is the coupling constant for the corresponding transition. Below, we assume that the modes $b$ and $c$ are strong classical fields, but here we need their quantum versions for writing rigorously the Schr\"odinger picture Hamiltonian, which has to be time-independent.

\subsection{First and second interaction pictures}

We define the first interaction picture by introducing the evolution operator $\mathcal{U}_{I}(t) = \mathcal{U}_0^\dagger(t)\mathcal{U}(t)$, where $\mathcal{U}_0(t)=\exp\left(-iH_0t\right)$. This operator satisfies the equation
\begin{equation}\label{SE1}
  i\frac{d}{dt}\mathcal{U}_I(t) = H_I \mathcal{U}_I(t),
\end{equation}
where
\begin{equation}\label{HI}
  H_I = \mathcal{U}_0^\dagger(t) H \mathcal{U}_0(t) - H_0 = ga\sigma_{31} + d_{12}b\sigma_{21} + d_{23}c\sigma_{32} + \textrm{H.c.}
\end{equation}
Now we assume that the modes $b$ and $c$ are strong classical fields and we can disregard their commutators, replacing the operators $b$ and $c$ by c-numbers $B$ and $C$ respectively, and defining the Rabi frequencies $2\Omega_{12}=2d_{12}B$ and $2\Omega_{23}=2d_{23}C$, which we assume to be real and positive.

Introducing the new atomic ground-state basis $|\pm\rangle = \left(|1\rangle\pm|2\rangle\right)/\sqrt{2}$, we rewrite Eq.~(\ref{HI}) as $H_I = H_1+V_1$, where
\begin{eqnarray}\label{H1}
  H_1 &=& \Omega_{12}\left(\sigma_{++}-\sigma_{--}\right),\\\label{V1}
  V_1 &=&   \frac{g}{\sqrt{2}}a\left(\sigma_{3+}+\sigma_{3-}\right) + \frac{\Omega_{23}}{\sqrt{2}}\left(\sigma_{3+}-\sigma_{3-}\right) + \textrm{H.c.}
\end{eqnarray}

The second interaction picture is defined by introducing the evolution operator $\mathcal{U}_{J}(t) = \mathcal{U}_1^\dagger(t)\mathcal{U}_I(t)$, where $\mathcal{U}_1(t)=\exp\left(-iH_1t\right)$. This operator satisfies the equation
\begin{equation}\label{SE2}
  i\frac{d}{dt}\mathcal{U}_J(t) = H_J(t) \mathcal{U}_J(t),
\end{equation}
where
\begin{equation}\label{HJ}
  H_J(t) = \mathcal{U}_1^\dagger(t) H_I \mathcal{U}_1(t) - H_1 = \frac{ga+\Omega_{23}}{\sqrt{2}}\sigma_{3+}e^{-i\Omega_{12}t} + \frac{ga-\Omega_{23}}{\sqrt{2}}\sigma_{3-}e^{i\Omega_{12}t} + \textrm{H.c.}
\end{equation}

\subsection{Effective Hamiltonian \label{sec:Heff}}

In the strong ground-state coupling regime, where $\Omega_{12}\gg(g,\Omega_{23})$, all terms of the interaction Hamiltonian, Eq.~(\ref{HJ}), are fast oscillating and the state of the system at times far exceeding the period of these oscillations $2\pi/\Omega_{12}$ can be approximately found by replacing $H_J(t)$ by an effective Hamiltonian $H_\mathrm{eff}$, which is found in the following way \cite{James07}.

We rewrite Eq.~(\ref{HJ}) in a harmonic form grouping together the positive and negative frequency parts, as $H_J(t)=he^{-i\Omega_{12}t} + h^\dagger e^{i\Omega_{12}t}$, where
\begin{equation}\label{h}
  h = \frac{ga+\Omega_{23}}{\sqrt{2}}\sigma_{3+} + \frac{ga^\dagger-\Omega_{23}}{\sqrt{2}}\sigma_{-3}.
\end{equation}
Now the effective Hamiltonian is
\begin{equation}\label{Heff}
  H_\mathrm{eff} = \frac{\left[h^\dagger,h\right]}{\Omega_{12}}  = H_2 + V_2 + V_2',
\end{equation}
where
\begin{eqnarray}\label{H2}
H_2 &=& \left(\delta a^\dagger a+r^2\delta\right) \left(\sigma_{++}-\sigma_{--}\right),\\\label{V2}
V_2 &=& r\delta\left(a+a^\dagger\right)\left(\sigma_{++}+\sigma_{--}\right),\\\label{V2prime}
V_2' &=& -2r\delta\left(a+a^\dagger\right)\sigma_{33},
\end{eqnarray}
with the shortcuts $\delta=g^2/2\Omega_{12}$, $r=\Omega_{23}/g$.

In terms of the effective Hamiltonian the evolution operator of the second interaction picture is $\mathcal{U}_J(t) = \exp\left(-iH_\mathrm{eff}t\right)$. We see that the effective Hamiltonian is diagonal in the basis $\left\{|+\rangle,|-\rangle,|3\rangle\right\}$. It means that if the population of some of these basis states is zero initially, it remains zero with the course of time, which is a characteristic trait of the phenomenon of EIT \cite{Mucke10}. In the following, we assume that the atom is prepared initially in a combination of its two ground states, and therefore, the operator $V_2'$ may be disregarded.

\subsection{Third interaction picture}

The third interaction picture is defined by introducing the evolution operator $\mathcal{U}_K(t) = \mathcal{U}_2^\dagger(t)\mathcal{U}_J(t)$, where $\mathcal{U}_2(t)=\exp\left(-iH_2t\right)$. This operator satisfies the equation
\begin{equation}\label{SE3}
  i\frac{d}{dt}\mathcal{U}_K(t) = H_K(t) \mathcal{U}_K(t),
\end{equation}
where
\begin{equation}\label{HK}
  H_K(t) = \mathcal{U}_2^\dagger(t) H_\mathrm{eff} \mathcal{U}_2(t) - H_2 = r\delta\left(ae^{-i\delta t} + a^\dagger e^{i\delta t} \right) \sigma_{++} + r\delta\left(ae^{i\delta t} + a^\dagger e^{-i\delta t} \right) \sigma_{--}.
\end{equation}

The solution of Eq.~(\ref{SE3}) with the initial condition $\mathcal{U}_K(0)=\mathbf{1}$ can be written in the form of Dyson time-ordered series \cite{Berestetskii82}
\begin{equation}\label{Texp}
  \mathcal{U}_K(t) = \mathcal{T} e^{-i\int_0^t H_K(t')dt'},
\end{equation}
where $\mathcal{T}$ is the time-ordering operator, placing the operators $H_K(t)$ with higher time to the left in the Taylor expansion of the exponential. For sufficiently small $r$ this time ordering can be disregarded, leading to the solution $\mathcal{\tilde{U}}_K(t) \approx e^{\Xi_1(t)}$, where
\begin{equation}\label{Xi1}
  \Xi_1(t) = r \left[a\left(e^{-i\delta t}-1\right) - a^\dagger\left(e^{i\delta t}-1\right)\right]\sigma_{++} - r \left[a\left(e^{i\delta t}-1\right) - a^\dagger\left(e^{-i\delta t}-1\right)\right]\sigma_{--}.
\end{equation}

Coming back to the first interaction picture we obtain the evolution operator
\begin{equation}\label{UI1}
  \mathcal{\tilde U}_I(t) = \mathcal{U}_1(t)\mathcal{U}_2(t)\mathcal{\tilde U}_K(t) = e^{-i\tilde\varphi(t)\left(\sigma_{++}-\sigma_{--}\right)}D\left(\alpha_+\sigma_{++}\right)D\left(\alpha_-\sigma_{--}\right) e^{-i\delta a^\dagger at\left(\sigma_{++}-\sigma_{--}\right)},
\end{equation}
where $\tilde\varphi(t)=(\Omega_{12}+r^2\delta)t$, $\alpha_\pm=\mp r\left(1-e^{\mp i\delta t}\right)$, and $D(x)=e^{xa^\dagger - x^\dagger a}$ is an extension of Glauber's displacement operator to the case of operator-valued argument. The rightmost factor in the right-hand side of Eq.~(\ref{UI1}) can be omitted if the initial state of the field is vacuum, which we assume in the following.

The evolution operator, Eq.~(\ref{UI1}), acting on the initial state $|+\rangle|\mathrm{vac}\rangle$, where $|\mathrm{vac}\rangle$ is the vacuum state of the field, produces a coherent state of the field with the amplitude $\alpha_+$, leaving the atomic state unchanged. Similarly, acting on the initial state $|-\rangle|\mathrm{vac}\rangle$, it produces a coherent state of the field with the amplitude $\alpha_-$, leaving the atomic state unchanged. If the atom is initially prepared in a superposition of the states $|+\rangle$ and $|-\rangle$, then the state of the field and atom becomes entangled. For instance, choosing the initial state $|1\rangle|\mathrm{vac}\rangle$, we obtain the state at moment $t$
\begin{equation}\label{Psi}
  |\Psi(t)\rangle = \frac12|1\rangle\left(e^{-i\tilde\varphi(t)}|\alpha_+\rangle + e^{i\tilde\varphi(t)}|\alpha_-\rangle\right) + \frac12|2\rangle\left(e^{-i\tilde\varphi(t)}|\alpha_+\rangle - e^{i\tilde\varphi(t)}|\alpha_-\rangle\right),
\end{equation}
where $|\alpha_\pm\rangle$ is a coherent state with the amplitude $\alpha_\pm$.

Detecting the atom in the state $|1\rangle$, one obtains the field in a superposition of two coherent states with different amplitudes. Unfortunately, the approximation of time-ordering lifting implies small $r$ and, consequently, small $\alpha_\pm$. Thus, the state, Eq.~(\ref{Psi}), belongs to the class of ``Schr\"odinger kitten'' states \cite{Paris99}, which differ from the Schr\"odinger cat states by having an amplitude so small that the two components are not macroscopically distinguishable.

At arbitrarily high values of $r$, the time-ordering effects should be taken into account in the solution, Eq.~(\ref{Texp}), which is the subject of the next section.

\section{Time-ordering effects in EIT-OAL}
\subsection{Magnus expansion}

The interaction Hamiltonian $H_K(t)$, Eq.~(\ref{HK}), is proportional to $r$, and at small value of this parameter can be treated as a small perturbation. The traditional approach to treating a small perturbation in a Hamiltonian system consists in limiting the consideration to one or several first terms of the Dyson expansion, Eq.~(\ref{Texp}). This approach leads to the standard time-dependent perturbation theory, having found numerous applications in the description of evolution of various quantum systems. However, this standard approach has an important limitation: the obtained evolution operator is non-unitary in general and, as a consequence, does not preserve the commutation relations of the system variables. An alternative representation of the evolution operator is the Magnus expansion \cite{Blanes09}, which results from a Taylor decomposition  of $\ln\mathcal{U}_K(t)$ in the small parameter $r$:

\begin{equation}\label{Magnus}
  \mathcal{U}_K(t) = \exp\Big(\Xi_1(t)+\Xi_2(t)+\Xi_3(t)+...\Big),
\end{equation}
where $\Xi_k$ is of order $r^k$. The three first terms of the Magnus expansion are
\begin{eqnarray}\label{term1}
  \Xi_1(t) &=& -i\int_0^t H_K(t_1)dt_1,\\\label{term2}
  \Xi_2(t) &=& -\frac12\int_0^tdt_1\int_0^{t_1}dt_2 \left[H_K(t_1),H_K(t_2)\right],\\\label{term3}
  \Xi_3(t) &=& \frac{i}6\int_0^tdt_1\int_0^{t_1}dt_2\int_0^{t_2}dt_3 \left\{\left[H_K(t_1),\left[H_K(t_2),H_K(t_3)\right]\right]\right.\\\nonumber
  && + \left.\left[\left[H_K(t_1),H_K(t_2)\right],H_K(t_3)\right]\right\}.
\end{eqnarray}
When the infinite series is limited to any finite number of initial terms, the approximate evolution operator remains unitary, which is a great advantage of the Magnus expansion over a more conventional Dyson expansion.

The first term of this expansion, $\Xi_1(t)$, has been found in the previous section. Substituting Eq.~(\ref{HK}) into Eq.~(\ref{term2}), applying the commutation relation $\left[a,a^\dagger\right]=1$, and performing the integration, we find
\begin{equation}\label{Xi2}
  \Xi_2(t) = i\Delta\varphi(t)\left(\sigma_{++}-\sigma_{--}\right),
\end{equation}
where $\Delta\varphi(t)=r^2\left(\delta t - \sin\delta t\right)$. We see that at interaction times small compared to the period $t_0=2\pi/\delta$, the second term of Magnus expansion is close to zero, but at larger times it may become significant, especially with growing $r$.

We find also that $\Xi_3(t)$ and higher order terms of the Magnus expansion equal zero identically, because the nested commutators of the Hamiltonian $H_K(t)$ vanish. Thus, the evolution operator of the third interaction picture is exactly $\mathcal{U}_K(t)=\exp\Big[\Xi_1(t)+\Xi_2(t)\Big]$.

Thus, we have demonstrated another advantage of the Magnus expansion: in the case of the interaction Hamiltonian having zero nested commutator of some order $k$, the Magnus series reduces to a finite sum of  $k-1$ terms and can be calculated exactly.

\subsection{Solution with time-ordering effects \label{sec:Solution}}
\begin{figure}[h!]
\centering\includegraphics[width=8cm]{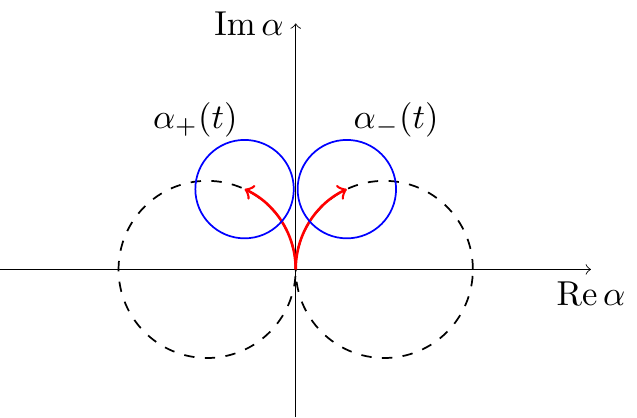}
\caption{Trajectories of the generated coherent states in the phase space of the cavity mode  (dashed lines) for $r=1.8$, and the trajectory traversed to the moment $t$ (red arrows). Blue circles show the two standard deviation areas of the Wigner functions of the coherent states. \label{fig:PhaseSpace}}
\end{figure}

Coming back to the first interaction picture we obtain the evolution operator with the time-ordering term
\begin{equation}\label{UIexact}
  \mathcal{U}_I(t) = \mathcal{U}_1(t)\mathcal{U}_2(t)\mathcal{U}_K(t) = e^{-i\varphi(t)\left(\sigma_{++}-\sigma_{--}\right)}D\left(\alpha_+\sigma_{++}\right)D\left(\alpha_-\sigma_{--}\right) e^{-i\delta a^\dagger at\left(\sigma_{++}-\sigma_{--}\right)},
\end{equation}
where $\varphi(t) = \tilde\varphi(t)-\Delta\varphi(t) = \Omega_{12}t+r^2\sin\delta t$. We see that the only effect of time-ordering is the change of the phase $\tilde\varphi(t)\rightarrow\varphi(t)$. As a result, the state of the atom and field, obtained from the initial state $|1\rangle|\mathrm{vac}\rangle$ after interaction time $t$ is given by Eq.~(\ref{Psi}) with the replacement $\tilde\varphi(t)\rightarrow\varphi(t)$. Detecting the atom at time $t$ in the state $|1\rangle$ or $|2\rangle$ projects the field onto the state $|\psi_+\rangle$ or $|\psi_-\rangle$ respectively, where
\begin{equation}\label{psipm}
  |\psi_\pm(t)\rangle = \frac{\mathcal{N}_\pm}{\sqrt{2}}\left(e^{-i\varphi(t)}|\alpha_+\rangle \pm e^{i\varphi(t)}|\alpha_-\rangle\right),
\end{equation}
with $\mathcal{N}_\pm$ being normalization coefficients guaranteeing $\langle\psi_\pm|\psi_\pm\rangle=1$. The trajectories of the generated coherent states in the phase space of the cavity mode are shown in Fig.~\ref{fig:PhaseSpace}. Each amplitude $\alpha_+(t)$ or $\alpha_-(t)$ traverses a circular trajectory of radius $r$ with a period $t_0=2\pi/\delta$. The relative phase between the quantum states $|\alpha_+\rangle$ and $|\alpha_-\rangle$, $\varphi(t)$ changes very fast, because in the considered regime $\Omega_{12}/\delta=2\Omega_{12}^2/g^2\gg1$.

At the beginning of the interaction, at $t=0$, $|\psi_+(0)\rangle=|\mathrm{vac}\rangle$, while $|\psi_-(0)\rangle$ does not exist, since the probability of detecting the atom in the state $|2\rangle$ is zero. We see from  Fig.~\ref{fig:PhaseSpace}, that with the course of interaction the amplitudes $\alpha_+$ and $\alpha_-$ diverge on the phase plane. If the circle radius $r$ is sufficiently large, at certain moment the Wigner functions of these states become well separated, and a Schr\"odinger cat state is formed. It happens approximately when the distance between the coherent states $\left|\alpha_+-\alpha_-\right|$ surpasses four standard deviations of the Wigner function of each coherent state, as shown in Fig.~\ref{fig:PhaseSpace}. We recall that the Wigner function of a coherent state $|\alpha_0\rangle$ is \cite{Cahill69b}
\begin{equation}\label{W}
  W(\alpha)=2\exp\left(-2\left|\alpha-\alpha_0\right|^2\right),
\end{equation}
and its standard deviation in any direction is $\sigma=\frac12$.

Since the relative phase $\varphi$ between the two coherent components changes very fast, the state $|\psi_+\rangle$ passes through all possible variants of the cat state: (displaced) even coherent state ($\varphi=0$), odd coherent state ($\varphi=\pi/2$), Yurke-Stoler coherent state ($\varphi=\pi/4$) many times during one period $t_0$ of moving on the circle on the phase plane. The same concerns the state $|\psi_-\rangle$ at different moments. The maximal separation of the coherent components is reached, obviously, at $t=t_0/2$, where $\alpha_\pm=\mp2r$. At this moment the cat state has the maximal size and, for sufficiently high $r$, the maximal nonclassicality.

To observe how a nonclassical cat state appears from the classical vacuum state, we use a nonclassicality measure, known as ``total noise'' \cite{Hillery89}, which is defined for arbitrary pure state $|\psi\rangle$ of an optical mode as
\begin{equation}\label{T}
  T(\psi) = \langle\psi|a^\dagger a|\psi\rangle - \left|\langle\psi|a|\psi\rangle\right|^2.
\end{equation}
This measure is zero for all coherent states, which are the only classical states among the pure states. For all other pure states $T(\psi)$ is positive and gives a good estimate of how nonclassical the state is. Total noise has important extensions to the class of mixed states of one optical mode, such the maximal quadrature quantum Fisher information \cite{Yadin18,Kwon19} having an explicit operational meaning, and the ordering sensitivity \cite{DeBievre19} directly related to the decoherence scale \cite{Hertz20}.

\begin{figure}[ht!]
\centering
\includegraphics[width=7cm]{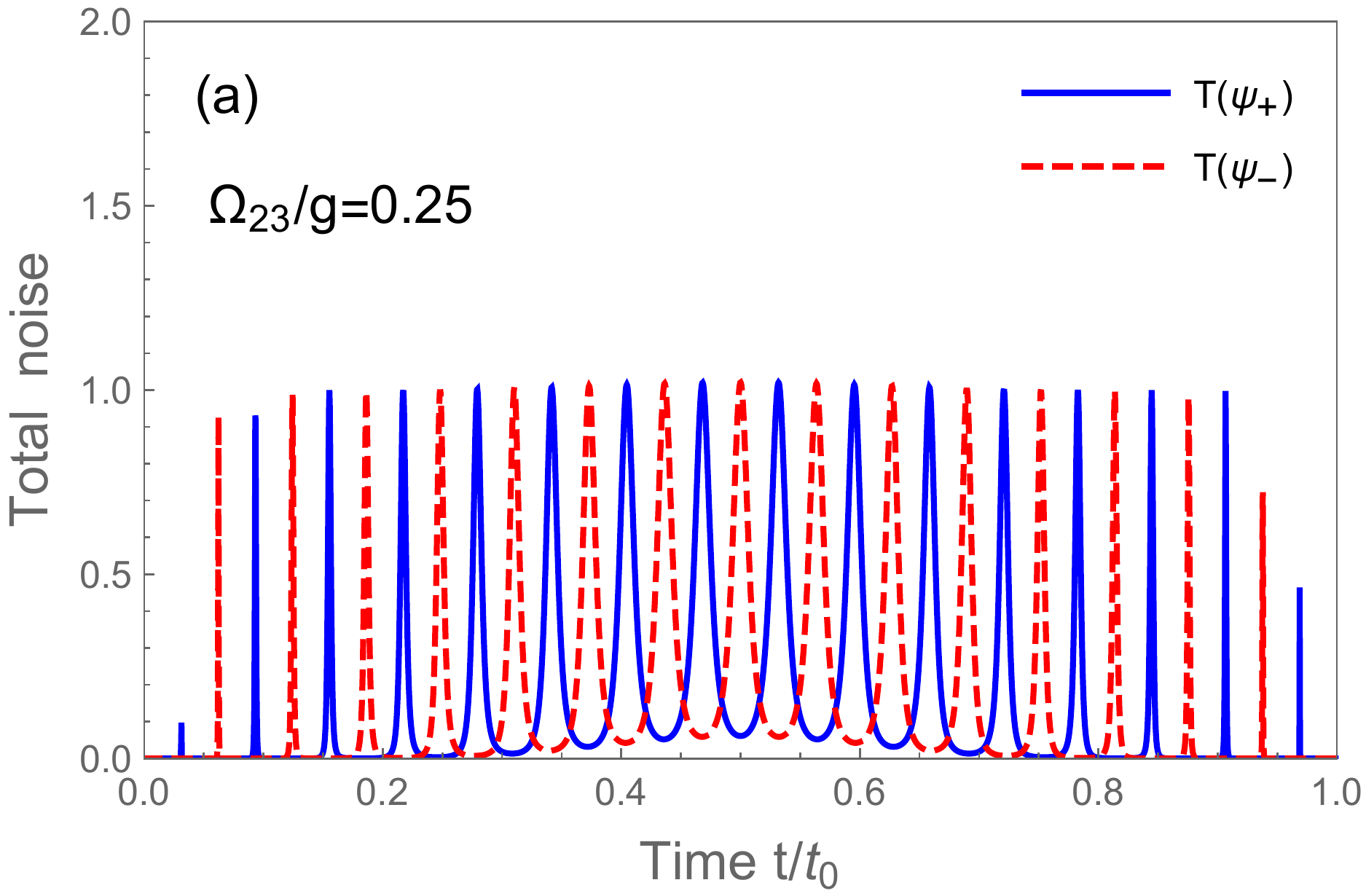}
\includegraphics[width=7cm]{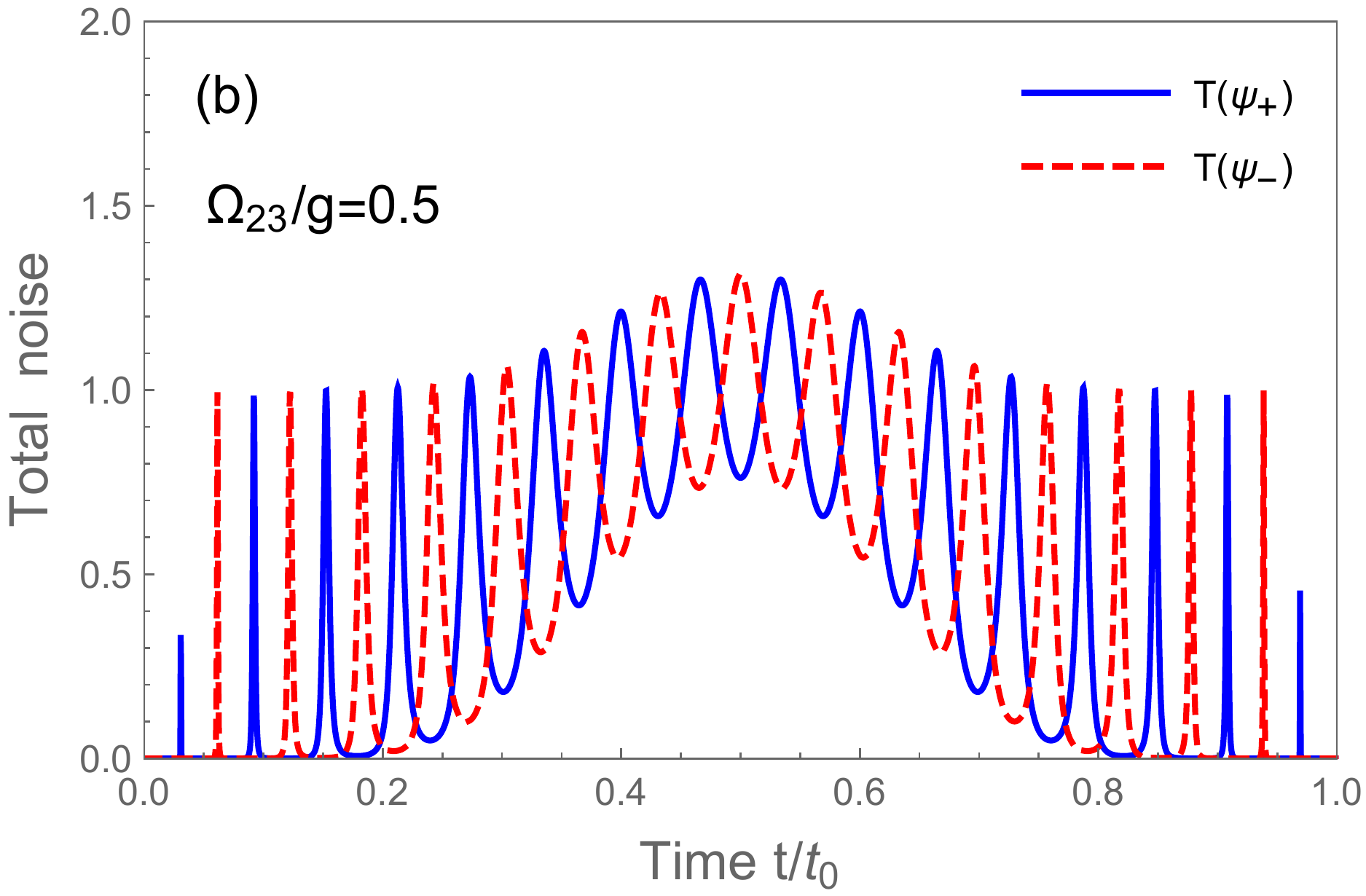}
\includegraphics[width=7cm]{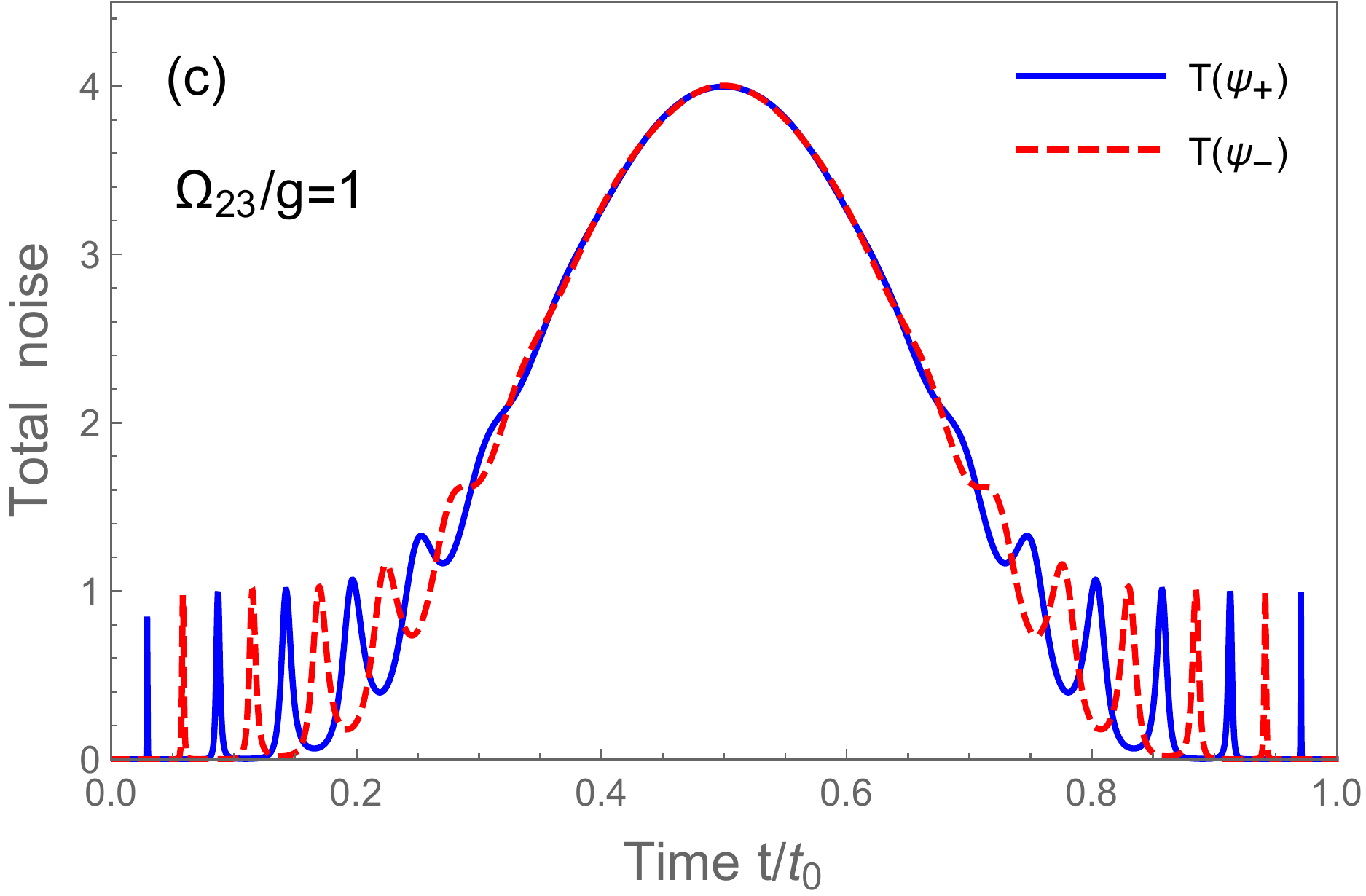}
\caption{Total noise of the conditional state of the cavity field under condition that the atom is detected in one of its ground states for different values of the parameter  $r=\Omega_{23}/g$. The ratio $\Omega_{12}/\delta$, which should be much greater than 1, is just 8 here for illustration. \label{fig:T}}
\end{figure}

Substituting Eq.~(\ref{psipm}) into Eq.~(\ref{T}) we obtain
\begin{equation}\label{Tpm}
  T(\psi_\pm) = \frac{\left|\alpha_+-\alpha_-\right|^2\left(1-e^{-\left|\alpha_+-\alpha_-\right|^2}\right)}{\left[2\pm q\pm q^*\right]^2},
\end{equation}
where
\begin{equation}\label{q}
  q=\langle\alpha_+|\alpha_-\rangle e^{2i\varphi} = e^{-\frac12|\alpha_+|^2-\frac12|\alpha_-|^2+\alpha_+^*\alpha_-+2i\varphi}.
\end{equation}
This function is shown in Fig.~\ref{fig:T} for various values of $r$. We see that at $r=0.25$, where the maximal separation of the coherent amplitudes, $4r=1$, equals two standard deviations of the coherent state Wigner function, the maximal value of the total noise is about 1. This value is reached at the relative phase of $\varphi=\pi/2$, i.e. for a displaced odd coherent state. The odd coherent state for the case of low amplitude is very close to the one-photon Fock state, for which the total noise is equal to 1. We note that the total noise is invariant with respect to displacement of the state on the phase plane. Thus, the total noise of a displaced odd coherent state is also close to unity. For other values of $\varphi$ the nonclassicality is close to zero, because the state has a dominant vacuum component. We see that the states $|\psi_+\rangle$ and $|\psi_-\rangle$ reach the maximal nonclassicality at different moments, because the relative phase of coherent components differs for these states by $\pi$. In this low-$r$ regime the main factor influencing the nonclassicality is the relative phase between the coherent components, while their separation is not important.

At a higer value of $r=0.5$ we see that a mesoscopic cat state is formed around $t=t_0/2$. This time corresponds to half-period of the motion along the trajectory in the phase space (dashed circle in Fig.~\ref{fig:PhaseSpace}), when the separation of the coherent amplitudes is maximal and equal to $4r$. For the considered value of $r=0.5$ the separation is $4r=2$, corresponding to 4 standard deviations of the coherent state Wigner function. Total noise is still oscillating, having higher values for a displaced odd coherent state than for other states, but there is also a distinct dependence on the separation of the coherent components. At even higher value of $r=1$ the separation of components becomes the main factor influencing the nonclassicality. At times close to $t=t_0/2$ a cat state is formed and the relative phase plays no role in the state nonclassicality.

Analyzing the evolution of the field state we need to remember that the applicability of the effective Hamiltonian approach, introduced in Sec.~\ref{sec:Heff}, requires an interaction time much higher than $2\pi/\Omega_{12}$. It means that the conditional state of the cavity field is given by Eq.~(\ref{psipm}) after the relative phase has made many periods of its evolution. Consequently, the initial oscillations in Fig.~\ref{fig:T} should be disregarded. It should be noted that the frequency of oscillations shown in this figure is rather low for the illustration purpose. Applicability of the condition $\Omega_{12}\gg(g,\Omega_{23})$ of the strong ground-state coupling regime requires that the oscillation frequency is at least 5 times higher.

\subsection{Effect of time ordering}

It is interesting and instructive to observe the effect of time ordering on the generated state and to determine the minimal value of $r$ where it becomes important. Total noise is not the best tool for this purpose because, as we saw in Sec.~\ref{sec:Solution}, the net effect of the time ordering is a change of the relative phase between the two coherent components, while the total noise becomes insensitive to this phase at a good separation of the components. To observe the effect of time ordering, we consider the task of generation of odd and even coherent states of light, having a paramount importance for optical quantum computation \cite{Guillaud19}. For this purpose we calculate the average parity of state
\begin{equation}\label{P}
  P(\psi) = \langle\psi|(-1)^{a^\dagger a}|\psi\rangle.
\end{equation}
The odd coherent state $|\psi_\mathrm{o}\rangle=\left(|\alpha\rangle-|-\alpha\rangle\right)[2(1-e^{-2|\alpha|^2})]^{-1/2}$ has only odd number states in its Fock-state decomposition, and its average parity is $P(\psi_\mathrm{o})=-1$. Similarly, the even coherent state $|\psi_\mathrm{e}\rangle=\left(|\alpha\rangle+|-\alpha\rangle\right)[2(1+e^{-2|\alpha|^2})]^{-1/2}$ has only even number states in its Fock-state decomposition, and its average parity is $P(\psi_\mathrm{e})=1$. For other values of the relative phase the average parity lies between -1 and 1.

\begin{figure}[ht!]
\centering
\includegraphics[width=7cm]{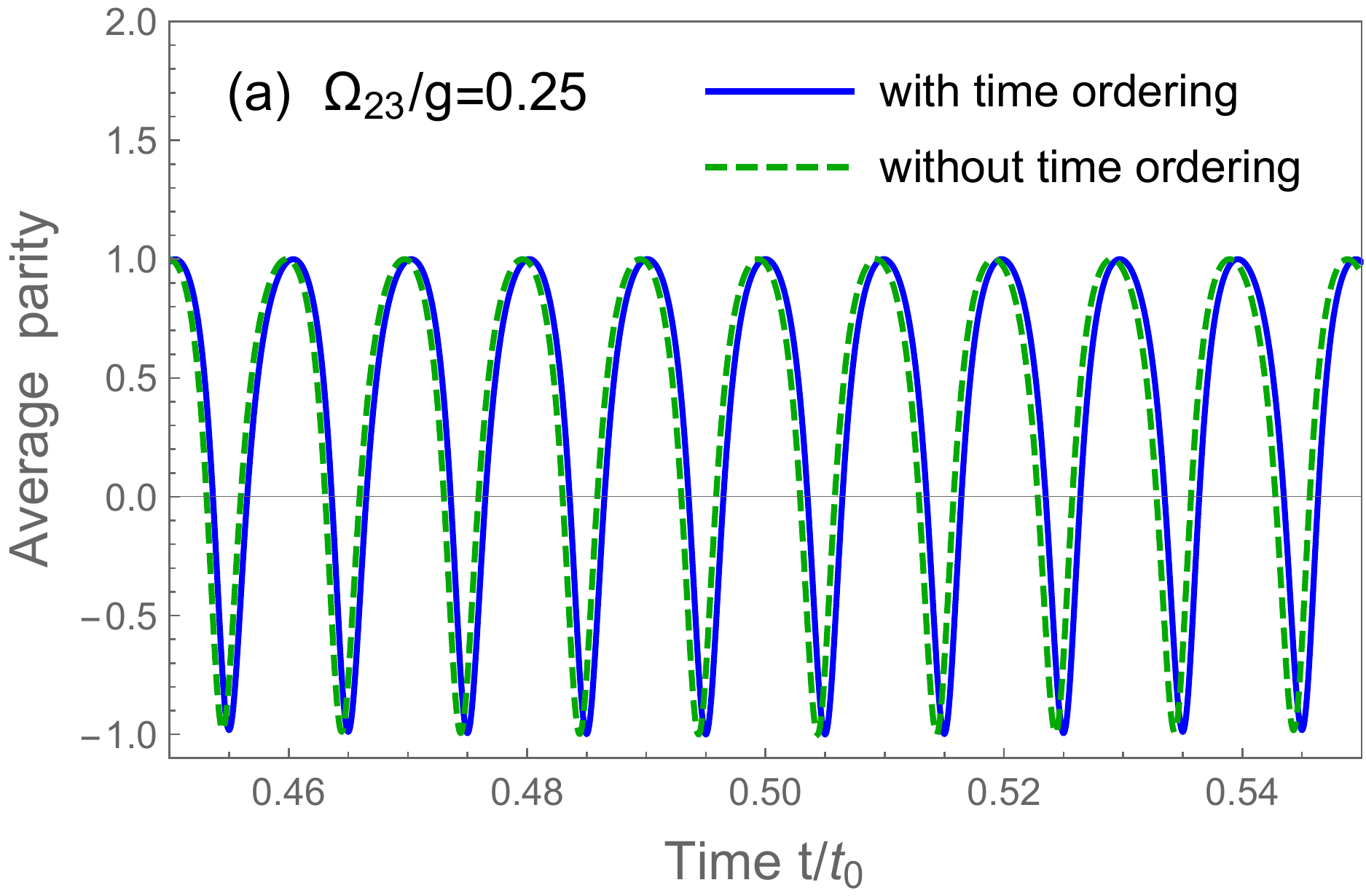}
\includegraphics[width=7cm]{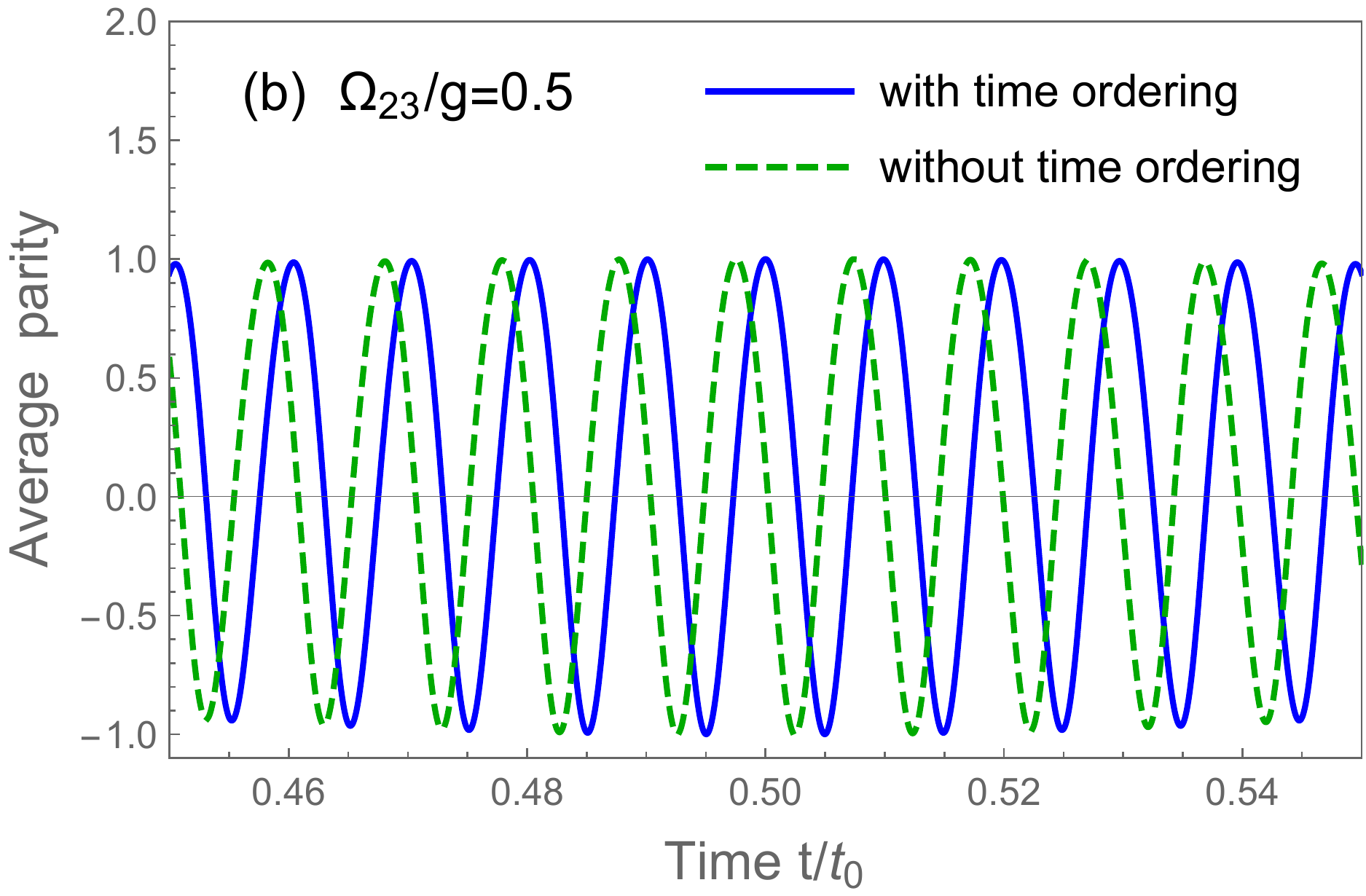}
\caption{Comparison of the average parity of the cavity field state $|\psi_+\rangle$ calculated with and without time ordering. The ratio $\Omega_{12}/\delta=50$. Average parity of 1 corresponds to generation of even coherent states, while that of -1 corresponds to generation of odd ones. \label{fig:ComparisonP}}
\end{figure}

The average parity of the state $|\psi_+\rangle$, generated at time $t$ is shown in Fig.~\ref{fig:ComparisonP}. We consider only a time interval around $t=t_0/2$, where the state amplitude is maximal. We see that at $r=0.25$ the effect of time ordering on the time of odd (or even) state generation is negligible, while at $r=0.5$ it should be taken into account.

We can determine a critical value of $r_c$ above which the time ordering becomes important by the following argument. The period of the average parity corresponds to the phase $\varphi(t)$ growing by $\pi$. The phase $\Delta\varphi(t)$, appearing because of time ordering, has a value of $\pi r^2$ around the time $t=t_0/2$, most interesting for applications. We assume that the time ordering is non-negligible if this correction is at least one tenth of the period. Thus we arrive at the critical value $r_c=\sqrt{0.1}=0.32$. At this value of the trajectory radius, odd and even coherent states are generated around $t=t_0/2$ with the amplitude $2r_c$. It is interesting to note, that the average number of photons in the odd coherent state with the amplitude $2r_c$ is  $\langle n\rangle_\mathrm{o}=1.05$. This is much less than the value of $\langle n\rangle\approx4$, at which time ordering becomes important in single-pass generation of squeezed states of light \cite{Christ13,Lipfert18}. The even coherent state of this amplitude has even lesser average number of photons $\langle n\rangle_\mathrm{e}=0.15$. However, this state hardly has a practical value, having a dominant vacuum component.

We ascribe the influence of time ordering on the field state at that low average photon number to the strong-coupling regime of the atom-field interaction, where the field and the atom exchange an excitation many times within the interaction time without losing photons to the environment. This situation would correspond to the very long crystal limit in the single-pass parametric downconversion, where the pump and the signal fields would exchange photons many times. In this limit the time ordering may become important even at moderate squeezing.

\begin{figure}[ht!]
\centering
\includegraphics[width=7cm]{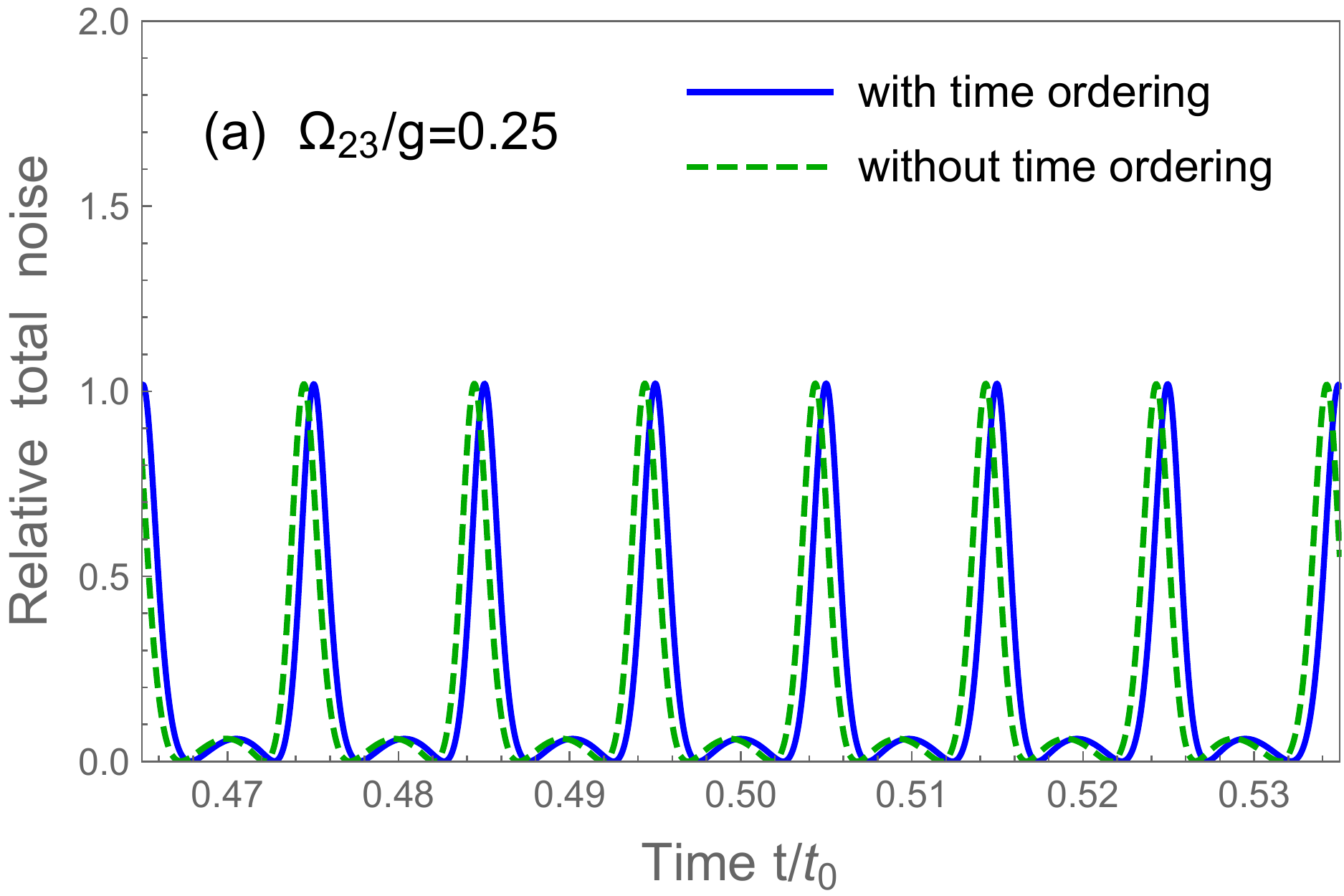}
\includegraphics[width=7cm]{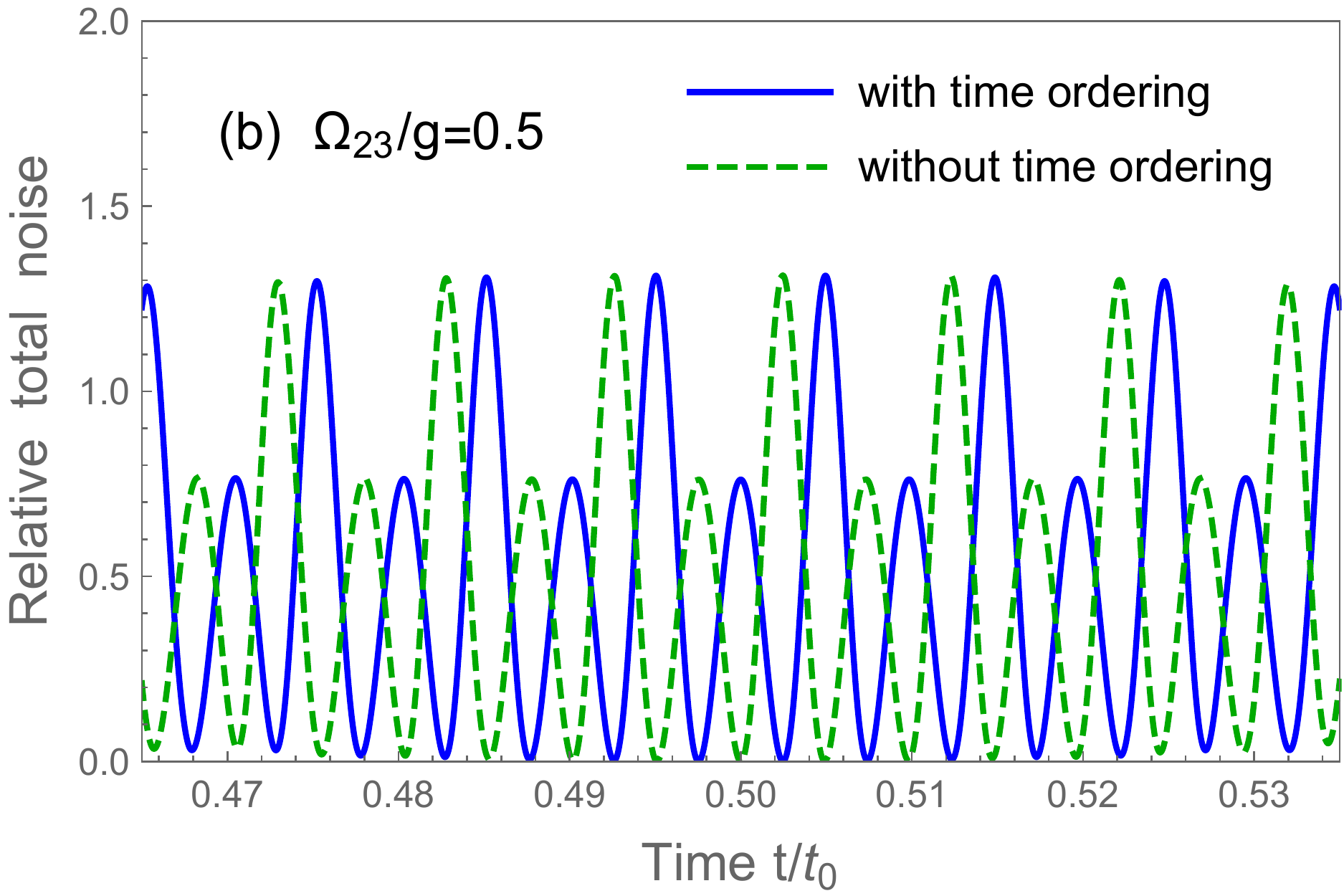}
\caption{Comparison of the relative total noise of the cavity field state $|\psi_+\rangle$ calculated with and without time ordering. The ratio $\Omega_{12}/\delta=50$. The relative total noise vanishes when a Yurke-Stoler coherent state is generated. \label{fig:ComparisonTA}}
\end{figure}

Another state, potentially interesting for applications, is the Yurke-Stoler coherent state \cite{Yurke86} $|\psi_\mathrm{YS\pm}\rangle = (|\alpha\rangle\pm i|-\alpha\rangle)2^{-1/2}$. This state is an eigenstate of the modified photon annihilation operator $A=e^{i\pi a^\dagger a}a$ \cite{Horoshko97jmo} and its decoherence can be slowed down by monitoring the cavity output light by photon counting and changing the phase after a count by a feedback loop \cite{Horoshko97prl,Horoshko98}. Thus, these states may be a good test bed for studying decoherence of the cat states. To determine the times of formation of the Yurke-Stoler coherent states, we introduce a measure similar to the total noise, but with the operator $a$ replaced by $A$:
\begin{equation}\label{TA}
  T_A(\psi) = \langle\psi|A^\dagger A|\psi\rangle - \left|\langle\psi|A|\psi\rangle\right|^2.
\end{equation}
This quantity, which we call ``relative total noise'', is vanishing for a Yurke-Stoler coherent state, similar to the vanishing of total noise for an ordinary coherent state. Thus, $T_A(\psi)=0$ means a creation of a state of our interest. The dependence of $T_A(\psi_+)$ on time is shown in Fig.~\ref{fig:ComparisonTA}.

As in the case of even and odd coherence states, we see that time ordering is negligible at low $r$, but becomes important at $r=0.5$. The critical value of $r$, above which time ordering effect should be calculated, is the same by the same argument. The average number of photons, corresponding to a Yurke-Stoler coherent state with the amplitude $2r_c$ is $\langle n\rangle_\mathrm{YS}=0.4$. Thus, we see again that time ordering becomes important in a cavity configuration at very low average photon number.

\section{Conclusions}

We have obtained the exact solution of the Schr\"odinger equation with a time-dependent effective Hamiltonian for the EIT-OAL in the regime of strong ground-state coupling. The net effect of the time-ordering correction has turned out to be in a modification of the relative phase between the two coherent components of an optical Schr\"odinger cat state generated in this regime. We have analyzed the parameters, corresponding to a non-negligible shift of the times of generation of definite variants of the cat state, as the odd and even coherent states and the Yurke-Stoler coherent state. We have found, that time-ordering becomes important at the average photon number in the cavity below 1, which is in striking contrast to the case of single-pass parametric downconversion, where it becomes important at average photon number in one optical mode above 4. We attribute this effect to multiple excitation exchanges taking place between the field and atom during the interaction time in the strong coupling regime. In this regime, inclusion of time-ordering terms in the evolution operator is necessary for the correct treatment of the stimulated emission.

These results shed new light on the importance of time-ordering for weak optical fields and may find their applications in the development of OALs and generation of nonclassical states of light for the tasks of quantum metrology and quantum information processing.

\section*{Funding}
D.B.H. and S.Y.K. are supported by Belarusian Republican Foundation for Fundamental Research, grant F20KI-035. C.-S.Y. is supported by the National Natural Science Foundation of China under Grant No.12011530014.


\section*{Disclosures}
The authors declare no conflicts of interest.

\section*{Data availability} All data in Figs.~3-5 are obtained from explicit analytical formulas given in the text of the article. No other data were generated or analyzed in the presented research.

\bibliography{BiblioOAL-2021}
\end{document}